\documentclass[letter]{./aa}
\usepackage{graphics}
\usepackage{txfonts}
\usepackage{natbib}
\bibpunct{(}{)}{;}{a}{}{,}

\DeclareRobustCommand{\ion}[2]{%
\relax\ifmmode
\ifx\testbx\f@series
{\mathbf{#1\,\mathsc{#2}}}\else
{\mathrm{#1\,\mathsc{#2}}}\fi
\else\textup{#1\,{\mdseries\textsc{#2}}}%
\fi}

\def\apjl{ApJ}%
\def\apjs{ApJS}%
\def\ao{Appl.~Opt.}%
\def\aap{A\&A}%
\begin{document}

\title{Opposite magnetic polarity of two photospheric lines in single spectrum 
 of the quiet Sun}
\author{R. Rezaei, R. Schlichenmaier, W. Schmidt, and O. Steiner}
\institute{Kiepenheuer-Institut f\"ur Sonnenphysik, 
Sch\"oneckstr. 6, 79104 Freiburg, Germany \\
E-mail: [rrezaei, schliche, wolfgang, steiner]@kis.uni-freiburg.de
}
\date{Received 3 March 2007/Accepted 24 April 2007}

\titlerunning{Opposite polarity in single spectrum}
\authorrunning{R.~Rezaei et al.}

\abstract
{}
{We study the structure of the photospheric magnetic field of the quiet Sun by 
investigating weak spectro-polarimetric signals.}
{ We took a sequence of Stokes spectra of the \ion{Fe}{i} 630.15\,nm and 630.25\,nm lines 
in a region of quiet Sun
near the disk center, using the POLIS spectro-polarimeter at the German VTT on Tenerife. 
The line cores of these two lines form at different heights in the atmosphere.
The 3$\sigma$ noise level of the data is about 1.8 $\times$ 10$^{-3}\,I_{\mathrm c}$. 
}
{ We present co-temporal and co-spatial Stokes-$V$ profiles of the \ion{Fe}{i}\,630\,nm line pair, 
where the two lines show opposite polarities in a single spectrum. 
We compute synthetic line profiles and reproduce these spectra with a two-component 
model atmosphere: a non-magnetic component and a magnetic component. 
The magnetic component consists of two magnetic layers with opposite polarity:
the upper one moves upwards while the lower one moves downward. 
In-between, there is a region of enhanced temperature.}
{The Stokes-$V$ line pair of opposite polarity in a single spectrum  
 can be understood as a magnetic reconnection event in the solar photosphere. 
We demonstrate that such a scenario is realistic, but the solution may not be unique.}

\maketitle

\section{Introduction}

Stokes polarimetry provides detailed information about the magnetic field  and its interaction 
with the plasma in the solar photosphere and chromosphere~\citep{stenflo_94,landi_landolfi}. 
While most of the observed Stokes-$V$ profiles in active regions and the network 
are close to antisymmetric with a low degree of asymmetry, abnormal and strongly asymmetric $V$ profiles are 
common in the inter-network~\citep{sigwarth01,Lites_02}. 
The classification of \cite{sigwarth_etal_99} presents abnormal $V$ profiles 
with a single lobe, two lobes with identical polarities, and 
$Q$-like, or pathological profiles with four or more lobes. 
There are indications from observations and 3-D simulations that the 
degree of asymmetry and the fraction of abnormal $V$ profiles 
increase with decreasing magnetic flux~\citep{sigwarth_etal_99, khomenko_etal_05}. 
While most of the pathological profiles can be reconstructed with models consisting of two or 
more magnetic components, \cite{grossman_etal_00} and \cite{steiner_2000} have shown that 
one-component models can also account for a large variety of $V$ profiles provided that magnetic, 
velocity, and temperature gradients are large enough.

An opposite polarity (OP) Stokes-$V$ profile is a set of two $V$ profiles of two 
different spectral lines, recorded in a strictly co-temporal and co-spatial observation, 
that shows different polarities. 
\cite{almeida_cerdena_kneer03} report opposite polarity $V$ profiles 
between the visible~(630\,nm) and the infrared~(1.56\,$\mu$m) neutral iron lines observed at   
 the French-Italian solar telescope TH\'EMIS and the 
German Vacuum Tower Telescope (VTT), respectively. 
This work was questioned later by \cite{khomenko_etal_05}, who showed that different seeing 
conditions for the two data sets can spuriously produce OP profiles.  

In this paper, we present observations of a quiet Sun region close to the disk center with 
the POlarimetric Littrow Spectrograph~\citep[POLIS,][]{schmidt03,beck05b} that show  
a few OP Stokes-$V$ profiles. 
Each set of OP profiles consists of the two iron lines at $\lambda\lambda$ 630.15 and 630.25\,nm 
that are part of  a {\it single spectrum} recorded strictly simultaneously.
The formation heights of these two lines span different layers in 
the photosphere~\citep[][and references therein]{cabrera_etal_05,khomenko_collados}. 
We study one set of these profiles in detail and argue that it hints at 
a magnetic reconnection event in the solar photosphere.

\begin{figure*}
\sidecaption
\resizebox{10.8cm}{!}{
\includegraphics*{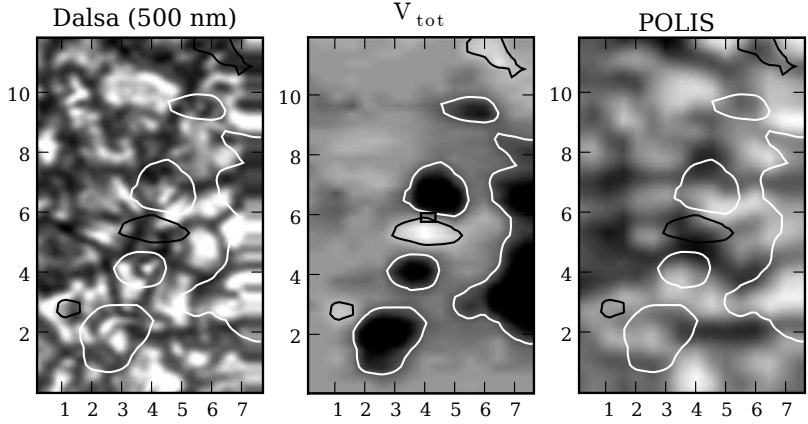}}
\caption[]{Co-spatial and co-temporal data of the speckle 
reconstructed continuum image at 500\,nm (left), $V_{\mathrm{tot}}$ map (middle), and 
Stokes-$I$ continuum map (right). Contours of the $V_{\mathrm{tot}}$ map are overplotted on all panels. 
The position of the opposite polarity profile is marked with a rectangle, which also shows 
the POLIS pixel size. The horizontal is the scan direction and the vertical is the slit direction. 
In the $V_{\mathrm{tot}}$ map, white and black indicate negative and positive 
polarities, respectively (for definition of positive/negative polarity, see Fig.~\ref{fig:op}). 
Tickmarks indicate arcsec.}
\label{fig:maps}
\end{figure*}

\section{Observations and data reduction}
A sequence of spectra taken in a quiet Sun region close to the disk center 
($\cos\theta$\,=\,0.99), was observed with the VTT in Tenerife, July 07, 2006. 
The seeing was good and stable during the observation. 
The Kiepenheuer Adaptive Optics System  was used for maximum spatial 
resolution and image stability~\citep{luhe_etal_03}. 
Each of the 38 scans consists of 16 slit positions. 
The scanning step size and spatial sampling along the slit were  0.48 and 0.29 arcsec, respectively.  
The scanning cadence was about 97~s. 

Full Stokes profiles of the neutral iron lines at 630.15 and 630.25\,nm 
and the Stokes-$I$ profile of the \ion{Ca}{ii}\,H line
were observed strictly simultaneously with the red (630\,nm) and blue (396.8\,nm)  channels of POLIS. 
The average continuum contrast (rms) of the POLIS Stokes-$I$ maps
is $\approx\,3.4\,\%$ of $I_{\mathrm{c}}$.
An absolute velocity calibration was performed using the telluric O$_2$ line 
at 630.20\,nm \citep{reza_etal_1}. The spectro-polarimetric data of the red channel 
were corrected for instrumental effects and telescope 
polarization with the procedures described by \cite{beck05b, beck05a}. 
The rms noise level of the Stokes parameters in the continuum  
is  $\sigma$\,=\,6.0\,$\times 10^{-4}$\,$I_{\mathrm{c}}$. 

Simultaneously, a continuum speckle channel in POLIS recorded a larger field of view at 500\,nm. 
The speckle reconstruction was performed using the Kiepenheuer-Institut Speckle 
Imaging Package~\citep{mikurda_luehe,fwoeger_phd}. 
The spatial resolution of the reconstructed image  is about 0.3~arcsec (cf. Fig.~\ref{fig:maps}, left panel). 
We used the POLIS intensity map and the reconstructed image to align the data.

We define the signed $V_{\mathrm{tot}}$ as follows:
\begin{equation}
V_{\mathrm{tot}} = \frac{\int^{\lambda_0}_{\lambda_{\mathrm b}} V(\lambda)\, \mathrm{d}\,\lambda  
- \int^{\lambda_{\mathrm r}}_{\lambda_0} V(\lambda)\, 
\mathrm{d}\,\lambda}{I_{\mathrm{c}}\,(\lambda_{\mathrm r} - \lambda_{\mathrm b})}{\rm ,}
\end{equation} 
where $\lambda_0$ is the zero-crossing wavelength of the Stokes-$V$ profile 
and $\lambda_{\mathrm r}$ and $\lambda_{\mathrm b}$ denote  fixed wavelengths in the red and blue 
continuum of the lines~\citep{lites_etal_99}. 
The continuum  speckle image and the POLIS Stokes-$I$ map are shown in the left 
and right panels of Fig.~\ref{fig:maps}, respectively. 
The middle panel of Fig.~\ref{fig:maps} shows the $V_{\mathrm{tot}}$ map in which   
the position of the OP profile is marked with a rectangle that also shows the POLIS pixel size. 
It is located between two patches of opposite polarities. 
The OP profile (exposure time 5\,s) was recorded within the time window of 15~s used for the speckle 
burst~(left panel, Fig.~\ref{fig:maps}). 
This image 
and the continuum map (right panel, Fig.~\ref{fig:maps}) indicate that the lower patch 
was co-spatial with an intergranular vertex (white patch in the $V_{\mathrm{tot}}$ map). 
Therefore, the $V$ profiles in the lower patch show a redshift, which is also 
the case for the 630.25\,nm OP profile.

Figure 2 shows the OP profile. The positive (negative) magnetic polarity 
in this figure corresponds to black (white) in the $V_{\mathrm{tot}}$ map.  
Both lines have positive amplitude and area asymmetries. 
The 630.15\,nm line of the OP profile has 
a blueshifted zero-crossing of $v_{\mathrm zc}\!\!\approx\!\!-1\,$km\,s$^{-1}$, while 
the 630.25\,nm line has a redshifted zero-crossing of $v_{\mathrm zc}\!\approx\!+2\,$km\,s$^{-1}$.

The OP profile along with its neighboring profiles are shown in Fig.~\ref{fig:array}. 
In this figure, the slit direction is vertical and the OP profile is at the center. 
The Stokes-$V$ profiles above and below the OP profile  
show strong $V$ signals of normal shape 
which shows that the OP profile was located at the center of a polarity reversal. 
The polarity of the 630.25\,nm line of the OP profile is the same as that of the lower pixels 
and the polarity of the 630.15\,nm line is identical to that of the upper ones. 
The $V$ profile left to the OP profile (Fig.~\ref{fig:array}, second row, left) is also strange: 
a pathological profile for the 630.15\,nm line and a regular profile of the 630.25\,nm line 
with a polarity similar to the 630.25\,nm line in the OP profile. 
The same is true for one scan-step after the OP profile (Fig.~\ref{fig:array}, second row, right column): 
a normal $V$ profile at 630.15\,nm  with a polarity identical 
to the 630.15\,nm line of the OP profile and no signal for the 630.25\,nm line. 

Note that the OP profile can not be reproduced by seeing effects as   
the adjacent profiles above and below the central OP profile in  
Fig.~\ref{fig:array} have incompatible shifts, line widths, and  
amplitudes: A least-squares fit to a superposition of the adjacent  
upper and lower profiles yields residuals
exceeding 2$\sigma$. Independently, we 
measure from the speckle burst that the standard deviation of the 
image motion during the 
exposure was only about 0.1\,arcsec.
Therefore the OP profile is not due to seeing effects. 

\begin{figure}
\begin{center}
\resizebox{8.9cm}{!}{\includegraphics*{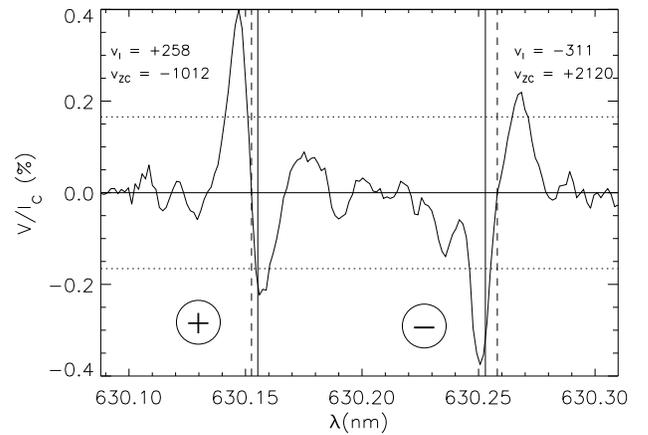}}
\caption[]{Stokes-$V$ line pair that shows opposite polarity in the 630.15 and 630.25\,nm lines. 
The position of the Stokes-$I$ line-core and the Stokes-$V$ zero-crossing are indicated by 
vertical solid and dashed lines, respectively. 
The line-core (v$_\mathrm{\begin{scriptsize}I\end{scriptsize}}$) 
and zero-crossing (v$_{\mathrm{\begin{scriptsize}zc\end{scriptsize}}}$) velocities are in m\,s$^{-1}$. 
}
\label{fig:op}
\end{center}
\end{figure}
\section{The model profile}
\begin{figure}
\resizebox{\hsize}{!}{\includegraphics*{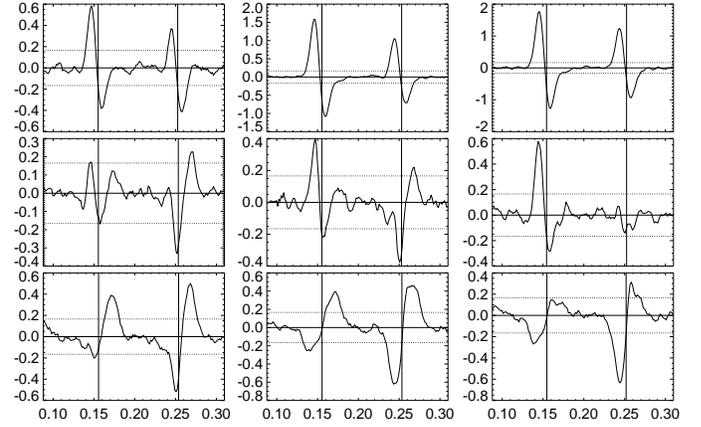}}
\caption[]{The OP Stokes-$V$ profile (center) and its surrounding profiles. 
Abscissae show the wavelength ($\lambda - 630$\,nm). The slit direction is vertical. 
Other parameters are like in Fig.~\ref{fig:op}.}
\label{fig:array}
\end{figure}

\begin{figure}
\centerline{\resizebox{8cm}{!}{\includegraphics*{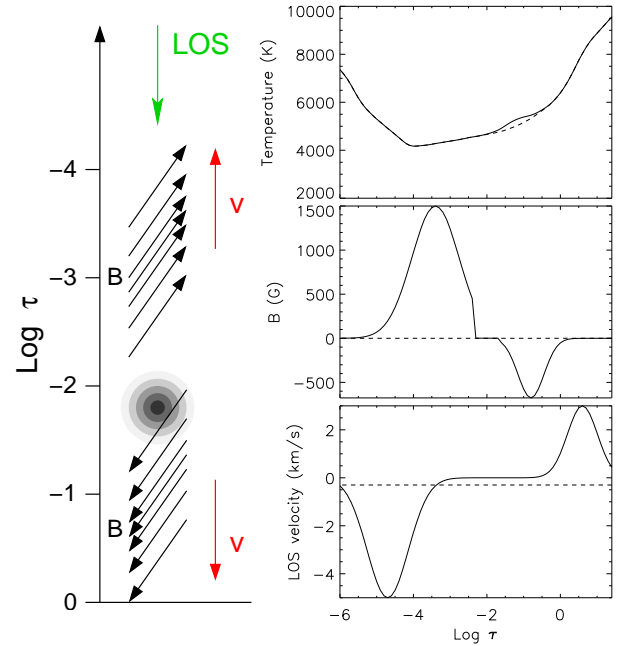}}}
\caption[]{Left: schematic configuration of the magnetic field, the line-of-sight velocity, 
and the temperature enhancement~(gray scale).  
Right: stratification of the atmosphere as a function of optical depth for the 
magnetic\,(solid) and non-magnetic\,(dashed) components.  
From top to bottom: temperature, magnetic field strength 
(with the the sign denoting the polarity), and 
line-of-sight velocity. The non-magnetic atmosphere has a convective blueshift of 
300\,m\,s$^{-1}$.}
\label{fig:model}
\end{figure}

\begin{figure}
\resizebox{\hsize}{!}{\includegraphics*{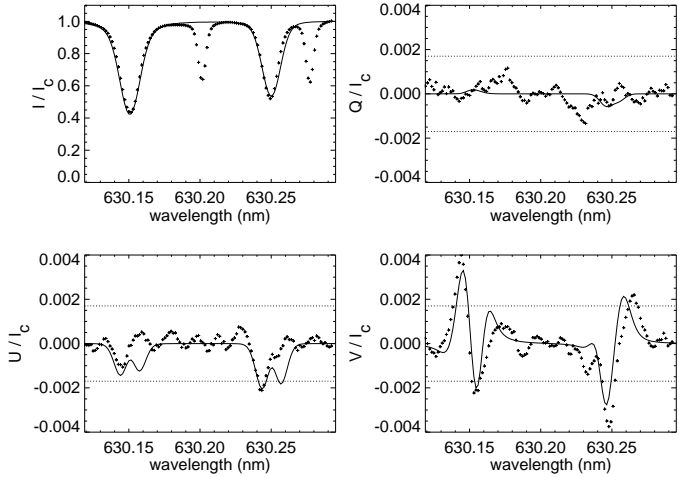}}
\caption[]{Comparison between the observed Stokes profiles (pluses) and the 
synthesized ones (solid lines). The horizontal dotted lines show the 3$\sigma$ noise level.}
\label{fig:fit}
\end{figure}

The basic properties of the observed OP profile are: (i) the opposite  
polarities of the $V$ profiles at 630.15\,nm and 630.25\,nm, (ii) the  
velocity of the deeper forming line (630.25\,nm) is positive 
(downward) and the velocity of the higher forming line  
(630.15\,nm) is negative (upward), and (iii) both  
lines show positive area and amplitude asymmetries. The vicinity of  
the OP profile is characterized by two patches of opposite polarity  
that are separated by only one resolution element as indicated by the  
rectangle in the middle panel of Fig.~\ref{fig:maps} and in the  
profile array of Fig.~\ref{fig:array}. This configuration suggests  
that we witness an electric current sheet in the intermediate  
atmospheric layers with magnetic field of one polarity in deeper  
layers and of the opposite polarity in higher layers. 
The field lines should be inclined with respect to the line-of-sight 
since there is a $U$ signal in the 
observed Stokes parameters~(Fig.~\ref{fig:fit}). The finding  
of upward velocity in the higher layer and downward velocity in the  
deeper layer presents evidence for magnetic reconnection, since   
such an event produces a bipolar jet. 
In the central region in-between the two magnetic  
polarities one expects an enhanced temperature due to Joules 
heating~(gray scale, Fig~\ref{fig:model}).  
This basic picture is sketched in the left panel of Fig.~\ref{fig:model}.
Based on these ideas, we construct in the following a quantitative model  
which allows us to compute a set of synthetic Stokes profiles for a  
direct comparison with the observed OP profile.

We use the SIR code~\citep{sir92} to synthesize the Stokes profiles of
the \ion{Fe}{i}\,630\,nm pair.  In order to fit the degree of
polarization and to account for straylight, the model atmosphere must
contain a field free (non-magnetic) component beside the magnetic
component. The non-magnetic component is taken to be the HSRA model
atmosphere~\citep{hsra}.  This thermal stratification is also used for
the magnetic component with a slight modification: in an intermediate
layer around $\log\tau_{500\,\mathrm{nm}}\!\approx\!-1.2$, we
introduce a temperature bulge.  At the peak of this hot bulge, it is
270\,K hotter than in the HSRA atmosphere.  The hot bulge is partly
field free and partly overlaps with the magnetic field of the deeper
layer.  Above and below this hot bulge, the velocity and magnetic
field have opposite orientations.  The ranges of non-vanishing
magnetic field strength and velocity overlap, but do not have their
peak values at the same optical depth.  Constant values for the
inclination of the magnetic field with respect to the vertical
(35\degr) and the azimuth (135\degr) are used to fit the ratios
between $Q$, $U$, and $V$.  The model atmosphere that was finally used
to reproduce the OP profile is shown in Fig.~\ref{fig:model}~(right).
All structure elements of this model atmosphere are necessary
ingredients for a successful reproduction of the OP profile.
In particular, the temperature bump is essential to reproduce the different V-profile
polarities when using only one line-of-sight along which two magnetic
polarities are present. 
This demonstrates that OP profiles can be synthesized with a
realistic model, but note that this solution may not be unique.

To obtain the filling factor of the magnetic atmosphere, 
we perform an inversion of the regular Stokes profiles  in the surrounding pixels of the OP profile.  
The inversion set up is similar to that of \cite{luis_beck05}, except that 
we assume a constant value for the stray light and allow for a linear gradient of the line-of-sight 
velocity in the magnetic component. 
From these inversions we obtain magnetic filling factors of 10\,--\,30\,\%, so 
we assume a filling factor of 20\,\% for our model atmosphere. 

The upper layer of magnetic field along with the bump of negative velocity leads  
to a very asymmetric $V$ profile for both lines. This profile has the polarity of the 
630.15\,nm line of the OP profile and shows a higher amplitude in 630.15\,nm than in 630.25\,nm. 
The lower-layer magnetic field of opposite polarity produces an almost 
antisymmetric $V$ profile at  630.25\,nm, but the profile at 630.15\,nm is asymmetric. 
The combination of these two components gives the final fit to the data 
that is shown in Fig.~\ref{fig:fit}. 
The quality of the fit is satisfactory considering that we only used one single line-of-sight 
for this complex topology. With a slight shift of the velocity peaks we are also able to reproduce 
the pathological profile to the left to the OP profile (Fig.~\ref{fig:array}).

\section{Discussion}

The magnetic area with polarity rendered in dark in the $V_{\mathrm{tot}}$ map
of Fig.~\ref{fig:maps} was present throughout the observing time 
of 64 minutes and was persistently 
bright in \ion{Ca}{ii}\,H, which suggests that it belonged to network
magnetic fields. 
%
On the contrary, the opposite polarity, white patch, visible in the 
center of the $V_{\mathrm{tot}}$ map, was a transient feature -- it assembled from 
diffuse flux of white polarity, concentrated and intensified to the white patch seen in 
Fig.~\ref{fig:maps}, before it rapidly weakened. Within the scanning cadence of
97~s it virtually disappeared, not without influencing the opposite polarity (black)
neighboring patches to the upper and lower sides, which weakened during this time.
Such events must frequently take place considering that the magnetic flux in the quiet 
network is permanently replaced by and in interaction with mixed polarity flux
\citep{schrij_title_etal_97}. When flux concentrations 
of opposite polarity collide, flux cancellation and magnetic field reconnection are likely 
to be involved \citep{zwaan87}. At the reconnection site we expect conversion from
magnetic to thermal energy and the formation of a bipolar jet: processes that are 
compatible with the temperature bulge in between the two layers of oppositely directed
velocities and magnetic fields of the atmosphere of Fig.~\ref{fig:model}  and with the observed flux 
cancellation and weakening.

For a check of compatibility with the thermal, kinetic, and magnetic 
energy fluxes required in a reconnection event, we first consider a 
control volume given by the size of one POLIS pixel of 
$A_p = \Delta_x\times\Delta_y = 350\,\mbox{km}\times 210\,\mbox{km}$ 
and the height range spanned by the FWHM of the temperature bulge 
of $h = 135$\,km, which encompasses a mass of $m = 1.3\times 10^{12}$\,kg. 
With a bulge peak of $\delta T = 270$\,K we obtain from Newton's
law of cooling a radiative heat loss from this volume of 
\begin{equation}
\dot{Q} = 16\, \sigma\, \kappa\, T^3\, \delta T\, m\, f = 9\times 10^{16}\,{\mathrm W,}
\end{equation}
where $\sigma$ is the Stefan-Boltzmann constant, 
$\kappa\!=\!1.1\times 10^{-2}\,\mbox{m}^2\,\mbox{kg}^{-1}$ the
opacity corresponding to the mean density of 
$1.3\!\times\!10^{-4}\,\mbox{kg}\,\mbox{m}^{-3}$, 
$T\!=\!5100$\,K the corresponding temperature of the unperturbed 
background atmosphere, and $f=0.2$ the filling factor.
This radiative energy flux should be of the same order of magnitude 
as the kinetic energy flux, 
\begin{equation}
F_\mathrm{kin} = (1/2)\,\rho\, v^3\, A_p\, f, 
\end{equation}
carried by the bipolar reconnection jets. Using the values of the
lower, heavier (downward moving) layer with 
$v_\mathrm{LOS} = 3\,\mbox{km\,s}^{-1}$ and 
$\rho = 3.0\times 10^{-4}\,\mbox{kg\,m}^{-3}$, and assuming
the velocity to be directed parallel to the magnetic field we obtain 
$F_\mathrm{kin} = 1\times 10^{17}$\,W. The kinetic energy of the upwards moving 
jet is negligible, because of the lower density. 
The kinetic and thermal energy fluxes need to be sustained by
a corresponding influx of magnetic energy of
\begin{equation}
F_\mathrm{mag} = v\, (B^2/2\mu_0)\, \Delta_x\, h\, \sqrt{f}. 
\end{equation}
With $B = 0.1\,\mbox{T}$ we obtain 
$F_\mathrm{mag} = 1\times 10^{17}\,\mbox{W}$ if the
influx velocity is $1.2\,\mbox{km\,s}^{-1}$, which is a reasonable 
value for the horizontal velocity of photospheric magnetic flux concentrations.

Photospheric magnetic reconnection has previously been studied by
\citet{litvinenko99} and \cite{takeuchi_shibata_01}. Since we observe a
flux intensification prior to flux canceling for the white polarity
patch of Fig.~\ref{fig:maps}, the scenario of \citet{takeuchi_shibata_01} consisting of
reconnection induced by convective intensification may apply to this event.

We would like to stress that the model presented here is not necessarily unique and that 
other model atmospheres may also reproduce the data. 
However, we expect any alternative model to show strong gradients in velocity and 
magnetic field strength.


\section{Conclusion}
We have discovered several sets of opposite polarity Stokes-$V$ profiles. 
Each set is a single spectrum of the   
630.15 and 630.25\,nm neutral iron lines pertaining to a single resolution element.
We use a model atmosphere consisting of a non-magnetic and one magnetic component 
for synthesizing the observed profiles. 
The magnetic component contains strong gradients along the line-of-sight in the magnetic field 
strength and the velocity. In particular, the higher and deeper layers have opposite 
magnetic polarity.
In accordance with the observed red- and blue-shifts of the spectral lines, 
the proposed model atmosphere contains a bipolar flow of material 
along the line-of-sight and it contains a temperature enhancement in-between.  
These atmospheric elements provide evidence of magnetic reconnection in the solar photosphere. 
Our estimated values for the thermal, kinetic, and magnetic energy fluxes are also 
compatible with a reconnection event.\vspace{-.1cm}

\begin{acknowledgements}
We wish to thank C. Beck and F. W\"oger for their 
assistance during observation and data reduction. 
We are grateful to J. Bruls, O.~von~der~L\"uhe, and E.~Khomenko for their useful comments. 
The POLIS instrument has been a joint development of the High  
Altitude Observatory (Boulder, USA) and the Kiepenheuer-Institut. 
Part of this work was supported by the Deutsche Forschungsgemeinschaft (SCHM 1168/8-1).\vspace{-.2cm}
\end{acknowledgements}

\bibliography{rezabib}        

\end{document}